\documentclass[12pt]{article}
\usepackage[cp1251]{inputenc}
\usepackage[russian]{babel}
\usepackage{hyperref}
\usepackage{graphicx}
\def\RE{R_{\mbox{Э}}}
\def\RSCH{R_{\mbox{Ш}}}

\begin{document}
\title{Космология Фридмана: горы реальные и потенциальные}
\author{С.В. Дворянинов, В.О. Соловьев}
%
%
%
\maketitle

В 2015 году исполнилось 100 лет общей теории относительности (ОТО).
Александр Александрович Фридман в 1922 году получил из 10 довольно сложных уравнений ОТО два уравнения, описывающие рождение, жизнь и смерть 
наблюдаемой Вселенной. Они стали фундаментом бурно развивающейся науки -- космологии. В математике часто оказывается, что одни и те же уравнения описывают совершенно 
разные задачи. В физике также часто используются модели разных явлений. Мы покажем, что уравнения Фридмана получаются не только в теории Эйнштейна, но и в простой задаче классической механики. Для понимания этой статьи достаточно знать законы Ньютона и закон сохранения энергии,  уметь дифференцировать здесь не обязательно.

\subsection*{Модель первая: катание с гор}
Представьте невысокую ледяную горку и хорошо скользящие санки (летом их можно заменить на скейт). Вы разгоняетесь, взлетаете на самую вершину горы, а потом по инерции скатываетесь с противоположного склона. Ваша скорость уменьшается при подъеме и увеличивается при спуске.  Это говорит нам и жизненный опыт, и формула закона сохранения энергии
$$
\frac{mv^2}{2}+U=E.
$$ 
Здесь первое слагаемое -- кинетическая энергия,  второе -- потенциальная. В правой части стоит полная механическая энергия. Во многих школьных задачах она считается сохраняющейся точно, в жизни, конечно, приблизительно, тем точнее, чем меньше трение. Если бы трения не было совсем, то санки никогда бы и не остановились. Потенциальную энергию силы тяжести обычно рассматривают как функцию высоты, но нарисуйте сечение горки вертикальной плоскостью, проходящей через траекторию саней, наложите на рисунок оси координат --  и высота станет функцией от горизонтальной координаты $x$, а значит, подобрав соответствующий масштаб, вы получите график потенциальной энергии: $U(x)=mgh(x)$.

И наоборот. Если дан график потенциальной энергии для некоторого тела, положение которого может быть задано одним числом, т.е. одной координатой, то не задумываясь о том, какой силе соответствует эта потенциальная энергия, мы можем отождествить движение тела с движением санок по реальным горам или ямам, имеющим изображенный на данном графике профиль. Закон сохранения энергии позволяет выразить кинетическую энергию одномерного движения как функцию координаты тела. Конечно, нужно также знать и полную энергию $E$, которую можно найти по начальным данным: координате и скорости. Масса тела тоже должна быть известна. По кинетической энергии можно определить скорость, правда с точностью до знака, так как направлений движения может быть два.    

\includegraphics[width=40mm,height=40mm]{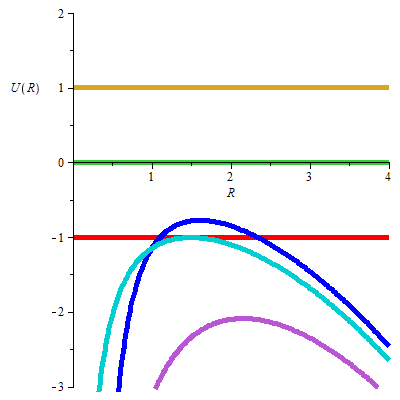}
Рис. 1

На Рис.1 показаны три кривые, изображающие график потенциальной энергии, заданной функцией от координаты $R$:
\begin{equation}
U(R)=-\frac{a}{R} -\frac{bR^2}{2},\label{eq:1}
\end{equation}
 при трех разных значениях параметра $b$, причем $a>0$, $b>0$. Конечно, кататься с такой горы можно только в  мыслях, с обеих сторон вы покатитесь в бездну.
Ниже мы покажем, каким физическим задачам может соответствовать  формула (\ref{eq:1}).

\includegraphics[width=8cm,height=10cm]{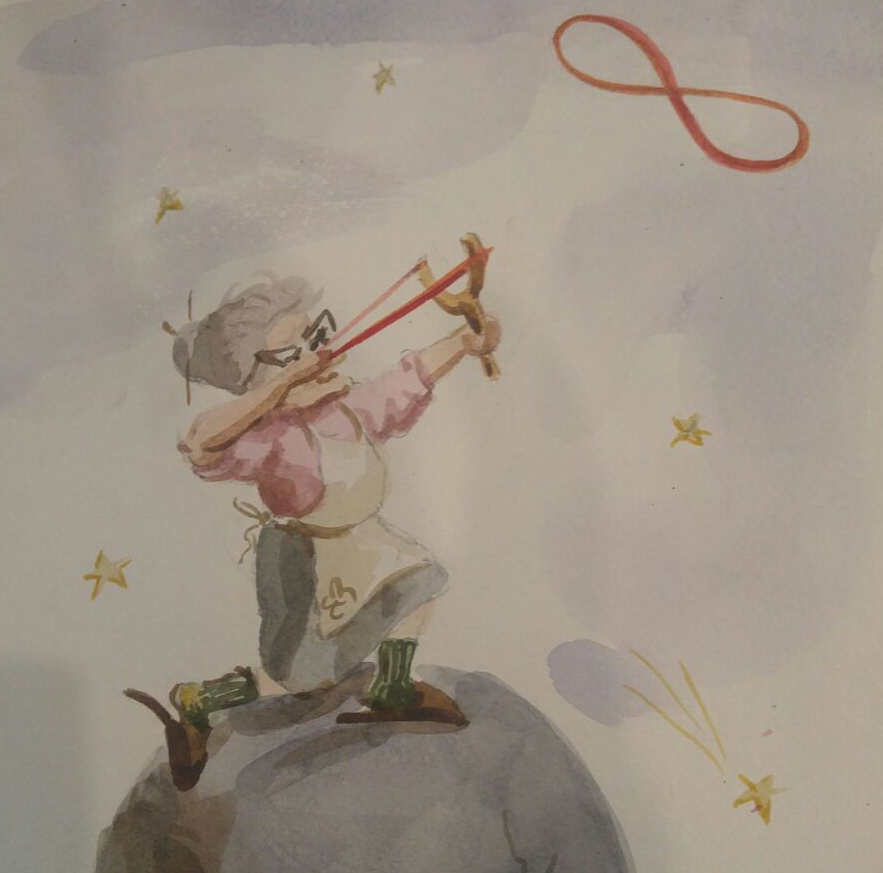}

Рис.2

\subsection*{Модель вторая: стрельба из рогатки в открытом космосе}
Сначала возьмем совсем простой случай $b=0$.  Пусть в космосе есть однородный шар массы $M$,  и 
с него вдоль направления его радиуса бросают пробное тело массы $m$, причем $m<<M$.  Тогда шар можно  считать неподвижным. Учтем закон всемирного тяготения в его полном виде, т.е. силу тяжести будем считать не постоянной, как вблизи поверхности Земли, а зависящей от расстояния между центрами тел
$$
F=-\frac{GMm}{R^2}.
$$ 
Тогда потенциальная энергия пробного тела будет
$$
U(R)=-\frac{GMm}{R},
$$
где $G=6,67\times 10^{-11}\ \frac{\mbox{м}^3}{\mbox{кг}\cdot\mbox{с}^{2}}$ -- гравитационная постоянная, а $R$ -- расстояние между центрами двух тел.
 Для тех, кто умеет дифференцировать, заметим, что силу можно найти, если известна потенциальная энергия $F=-U'(R)$. Как  будет двигаться подброшенное тело? Есть содержит ровно три возможности: при $E<0$ брошенное тело вернется и упадет на 
свое прежнее место, при $E=0$ оно не вернется, улетит бесконечно далеко и там остановится, наконец, последняя возможность -- при $E>0$ тело не остановится и на бесконечности. Во втором 
случае начальная скорость называется второй 
космической. Если тяжёлым телом будет Земля, а пробное тело бросают с её поверхности и пренебрегают сопротивлением атмосферы, то величина этой  скорости будет 
$$
v=\sqrt{\frac{2GM}{R_0}}=11 200\ \frac{\mbox{м}}{\mbox{с}},
$$ 
где $R_0\approx 6,38\cdot 10^{6}$ м -- радиус Земли, $M=5,98\cdot 10^{24}$ кг -- масса Земли.
Кстати, последняя формула, несмотря на то, что она была получена на основе механики Ньютона и без всякой связи с теорией относительности, может быть использована для нахождения гравитационного радиуса тела в ОТО. Достаточно подставить вместо $v$ скорость света $c$  и тогда
$$
\RSCH=\frac{2GM}{c^2},
$$
эта величина называется  радиусом Шварцшильда тела массы $M$. Если верить ОТО, то даже свет не может улететь от тела, радиус которого меньше, чем $\RSCH$. Читателю полезно вычислить этот радиус, например, для Земли или для Солнца.

\subsection*{Модель третья: Ньютон против анти-Гука}
Что может означать второе слагаемое в формуле (\ref{eq:1})?
Для $b<0$, оно положительно и совпадает с потенциальной энергией пружины. Воображаемая пружина тянет брошенное тело назад к центру шара по закону Гука $F=-kR$, причем $k=|b|$. Нас, однако, будет интересовать другой, весьма необычный случай: $b>0$. Тут пружина должна отталкивать пробное тело от центра шара, причем сила должна линейно возрастать с увеличением расстояния: $F=+bR$. В космологии за этим стоит так называемая тёмная энергия. В механике мы позволим себе называть это законом анти-Гука. Как наш художник представил себе эту персону вы увидите на Рис.3. Добавим также, что никто не испортил Ньютону больше крови, чем реальный Гук. Точку, в которой силы притяжения и отталкивания уравновешивают друг друга, уместно, как будет видно из дальнейшего, назвать точкой Эйнштейна. 

Теперь, наконец, выпишем здесь уравнения динамики Вселенной, полученные в 1922 г. А.А. Фридманом из общей теории относительности, сохраняя все его обозначения
\begin{eqnarray}
\frac{{R'}^2}{R^2}+\frac{2R''R}{R^2}+\frac{c^2}{R^2}-\lambda&=&0,\nonumber\\
\frac{3{R'}^2}{R^2}+\frac{3c^2}{R^2}-\lambda&=&\frac{\kappa}{2}c^2\rho.\label{eq:F}
\end{eqnarray}
Здесь $c\approx 3\cdot 10^8\ \frac{\mbox{м}}{\mbox{с}}$ -- скорость света, $\kappa=16\pi G/c^2$, $R$ -- ``радиус мира'' (или масштабный множитель), $\rho$ -- средняя плотность вещества во Вселенной, $\lambda$ -- космологическая постоянная, введенная А. Эйнштейном. Штрихи обозначают производные по времени, иными словами, $R'$ -- скорость, а $R''$ -- ускорение. Дальше мы постараемся убедить вас, что не решая, а только обдумывая эти уравнения и построив по ним несколько графиков можно понять закон эволюции Вселенной. 

Будем предполагать, вслед за Эйнштейном и Фридманом, что Вселенная   равномерно заполнена пылью,  не создающей давления, что пространство одинаково в любой точке и по любому направлению, имеет постоянную положительную кривизну и простейшую топологию, а значит, подобно трехмерной сфере. Не советуем мучиться и пытаться себе ее представить, представьте лучше себя двумерным существом живущим на  поверхности обычного шара.  Аналогом  площади двумерной сферы $4\pi R^2$ служит объем сферы трехмерной: $V_3=2\pi^2 R^3$. 
Ясно, что уравнения Фридмана можно преобразовывать, например, воспользовавшись вторым уравнением можно исключить из первого  $R'$. Можно, конечно, умножать и делить уравнения на выражения, отличные от нуля.  Давайте, вслед за Фридманом, введем полную массу вещества во Вселенной $M=\rho V_3=\rho 2\pi^2 R^3$. Эта величина сохраняется при изменении радиуса мира, что следует из наших уравнений, если применить дифференцирование. После выражения плотности через массу и радиус мира, переобозначения космологической постоянной в виде $\lambda=3\omega^2$
и умножения обеих частей на массу пробной частицы мы приходим к формулам
\begin{eqnarray}
ma&=&-\alpha\frac{ GmM}{R^2}+m\omega^2 R,\nonumber\\
\frac{m{v}^2}{2}&=&E-U(R),  \label{eq:2}
\end{eqnarray}
где для краткости использованы обозначения
$$
U(R)=-\alpha\frac{ GmM}{R}-\frac{m\omega^2 R^2}{2}, \qquad E=-\frac{mc^2}{2}. 
$$
Формулы (\ref{eq:2}) совпадают с уравнениями обсуждавшейся выше механической задачи о брошенном теле при  чуть измененной гравитационной постоянной $G\rightarrow \alpha G$, где $\alpha=V/V_3$ и при условии $\lambda=3\omega^2=0$. Множитель $\alpha=2/(3\pi)\approx 0.21$ возникает из-за замены выражения для объема шара в евклидовом пространстве $V=4/3\pi R^3$ на объём 3-мерной сферы $V_3=2\pi^2 R^3$. Появление $\alpha$ никак не отражается на качественной картине явления, которое мы собираемся описывать. Первая формула в (2) есть закон Ньютона, вторая -- закон сохранения механической энергии. Найдем координату точки Эйнштейна в модельной задаче (\ref{eq:2})
$$
R_{\mbox{Э}}=\sqrt[3]{
\frac{\alpha GM}{\omega^2},
}
$$
в космологии она будет играть важную роль.

Особенность ОТО состоит в том, что закон сохранения энергии оказывается одним из 10 уравнений ОТО, при наших предположениях оно превратилось во второе из уравнений Фридмана (\ref{eq:F}). Если в механике Ньютона для того, чтобы вывести из второго закона Ньютона закон сохранения энергии мы должны интегрировать по времени, то в ОТО интегрирования не требуется, все уже готово. Из-за этой причины энергия не вычисляется по начальным данным, а наоборот, на начальные данные накладывается ограничение, связь, эти данные оказываются связанными известным заранее значением энергии, в данном случае $E=-\frac12 mc^2$. Именно второе из уравнений (\ref{eq:F}) является главным, именно оно называется в научной литературе уравнением Фридмана. В суровые сталинские времена, когда философы, по выражению П.Л. Капицы, исполняли функции милиции, на уравнение ими был наклеен страшный ярлык ``Дифференциальное уравнение Бога''.

\subsection*{Динамика Вселенной}
Как должна вести себя Вселенная, подчиненная законам Фридмана? Так же как брошенное в радиальном направлении тело с потенциальной энергией (\ref{eq:1}), подчиненное законам Ньютона и анти-Гука. Обратите внимание, что эти законы не запрещают движение со скоростью сколь угодно большей скорости света.
Другая экзотика в том, что для лучшего сходства с эволюцией Вселенной нам придётся считать началом полёта тела ситуацию, когда расстояние $R$ почти равно нулю, скорость $v=R'$ почти равна бесконечности, но полная энергия конечна и равна $-\frac12 mc^2$ (однако, возможна иная точка зрения, см. последнюю часть этой статьи). Обратите внимание, что при такой постановке задачи тяжёлое, но точечное тело $M$ играет роль вещества Вселенной, а  движение пробного тела $m$ описывает изменение геометрии Вселенной. При этом динамика геометрии определяется количеством вещества.

Вы заметили, что согласно уравнениям (\ref{eq:2}) масса брошенного тела $m$ никак не скажется на его движении? Разделив второе из уравнений (2) на модуль полной энергии тела, то есть, на $mc^2/2$, и введя новую переменную $x=R/R_{\mbox{Э}}$, мы получим в обеих частях этого уравнения безразмерные величины
$$
\frac{v^2}{c^2}=-1+\frac{\beta}{3}\left(\frac{2}{x}+x^2\right),\qquad {\mbox{где}} \qquad \beta=\frac{\RSCH}{\pi\RE}.
$$

Строим графики безразмерной потенциальной энергии $\bar U=\frac{U}{|E|}$ при разных значениях параметра $\beta$:

\includegraphics[width=8cm,height=5cm]{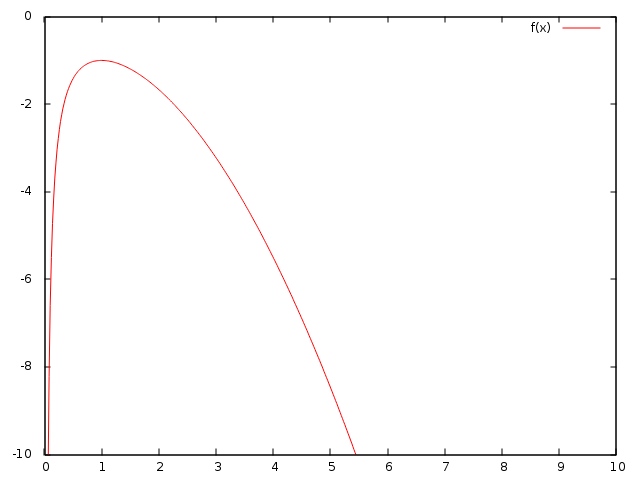}
 при $\beta=1$

\includegraphics[width=8cm,height=5cm]{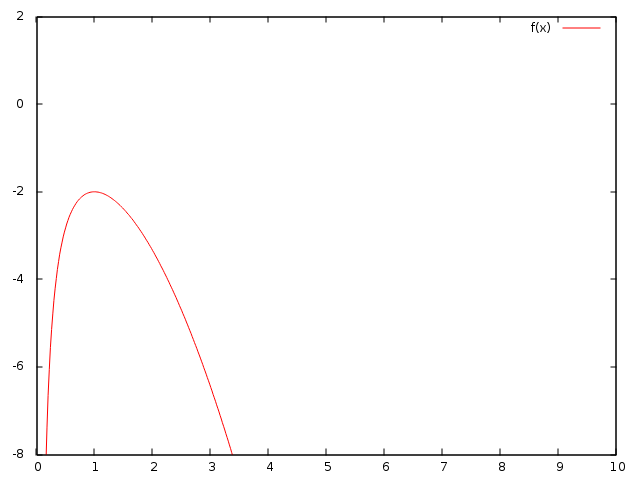}
 при $\beta=2$

\includegraphics[width=8cm,height=5cm]{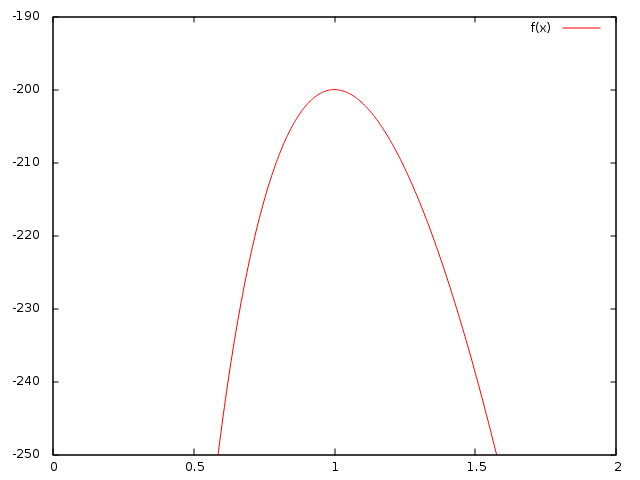}
 при $\beta=200$

Видно, что максимум потенциальной энергии всегда находится в точке Эйнштейна $x=1$. 

Построим теперь по отдельности потенциальные энергии Ньютона (красная линия), анти-Гука (зеленая линия) и их сумму (синяя линия).

\includegraphics[width=8cm,height=5cm]{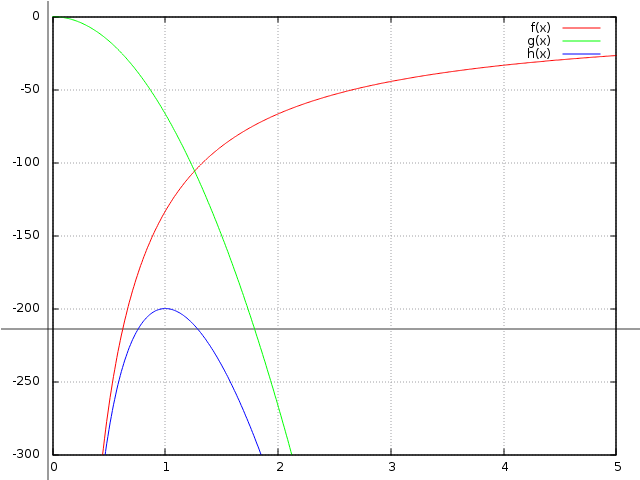}
 при $\beta=200$

Построим теперь графики  безразмерной скорости расширения Вселенной
$$
\frac{v}{c}=\sqrt{-1+\frac{\beta}{3}\left(\frac{2}{x}+x^2\right)}
$$

\includegraphics[width=8cm,height=5cm]{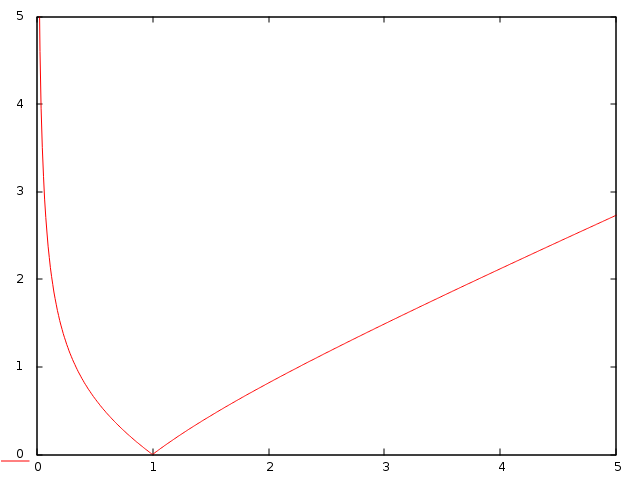}
 при $\beta=1$

\includegraphics[width=8cm,height=5cm]{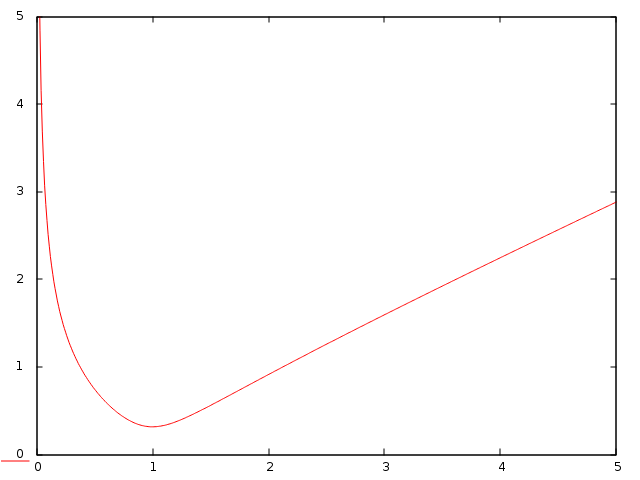}
 при $\beta=1.1$


\includegraphics[width=8cm,height=5cm]{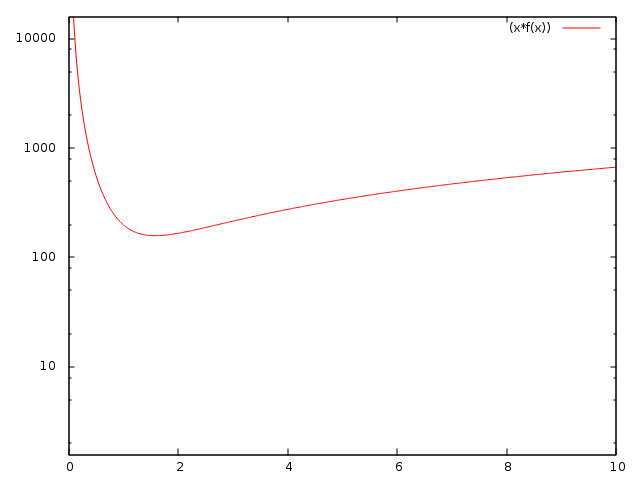}
 при $\beta=200$

Конечно, независимо от $\beta$, минимум  скорости всегда находится в точке Эйнштейна $R=\RE$, то есть, при $x=1$.

Построим графики для величины, которую назовём безразмерной постоянной Хаббла (из дальнейшего будет понятно, почему):

\begin{equation}
 \bar H\equiv\frac{v}{cx}=
\sqrt{-\frac{1}{x^2}+\frac{\beta}{3}\left(\frac{2}{x^3}+1\right)}\label{eq:H}
\end{equation}

\includegraphics[width=8cm,height=5cm]{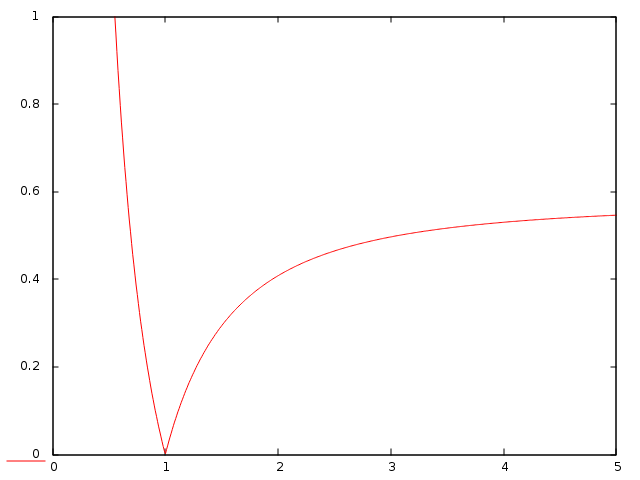}
 при $\beta=1$

\includegraphics[width=8cm,height=5cm]{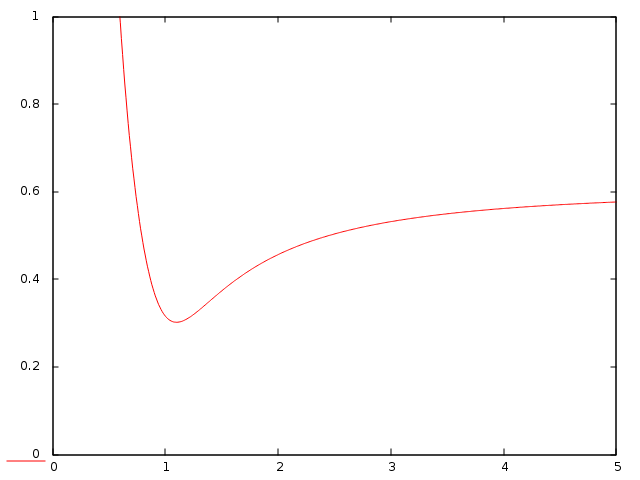}
 при $\beta=1.1$


\includegraphics[width=8cm,height=5cm]{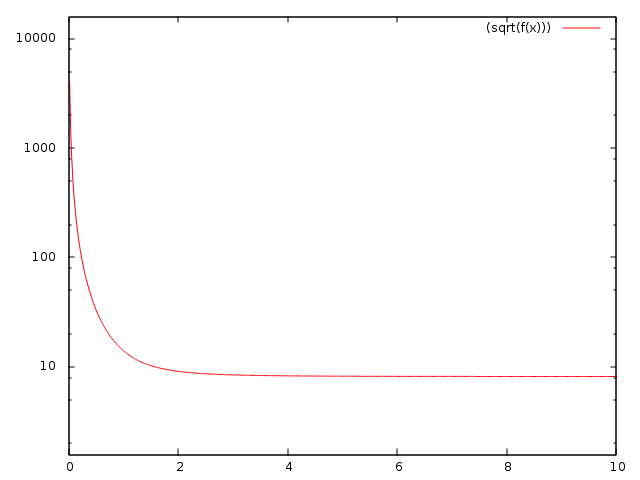}
 при $\beta=200$

Минимум здесь достигается при $x=\beta$, то есть, при $R=\RSCH/\pi$.

Вид графиков потенциальной энергии здесь, как и на Рис.1, напоминает крутую горку, по которой можно скатиться в бездну и слева, и справа.   Бездна соответствует бесконечно отрицательной потенциальной энергии, а значит, бесконечно большой кинетической, что видно из графиков скорости. Скатывание с горки справа налево соответствует все ускоряющемуся сжатию Вселенной, вплоть до нулевого радиуса, скатывание слева направо -- ускоренному расширению Вселенной, стремящейся к бесконечному радиусу. Достижима ли при движении слева направо для брошенного тела и для Вселенной вершина горы, она же -- максимум потенциальной энергии, она же -- точка Эйнштейна? Если она лежит выше значения полной энергии, то есть, $\bar U_{\mbox{max}}>-1$, то вершина недостижима. Это значит, что брошенное тело упадет обратно, а
Вселенная, начавшая в момент ``Большого Взрыва'' расширяться с бесконечной скоростью, будет постепенно замедляться, замрёт на мгновение и покатится обратно, наращивая скорость, обратно к нулевому радиусу --  сингулярности. Это происходит при $0<\beta<1$. Пройдёт ли Вселенная через сингулярность и начнёт ли расширяться снова, повторяя цикл за циклом? Фридман допускал такую возможность, вспоминая индусскую мифологию о периодах жизни. Он даже вычислил ``период мира'', взяв интеграл, принимая среднюю плотность вещества равной приведенной в книге английского астронома Эддингтона и считая космологическую постоянную нулевой. Получил 10 000 000 000 лет. Этот сценарий Фридман назвал ``периодическим миром''. Можно, конечно, думать, что падение нашего пробного тела приведёт к смерти Вселенной. Если точка Эйнштейна лежит ниже значения полной энергии, что происходит при $]beta>1$, то брошенное тело улетит в бесконечность, а Вселенная, соответственно, начав расширяться с бесконечной скоростью, будет сначала постепенно уменьшать скорость расширения, но затем, перевалив через вершину горы, снова начнёт ускоряться. Этот сценарий Фридман назвал ``монотонным миром первого рода''. Наконец, ``монотонный мир второго рода'' соответствует картине, когда тело начинает движение с нулевой начальной скоростью с правого склона горы и скатывается по нему с ускорением. ``Большого взрыва'' и горячей ранней вселенной в этом сценарии нет, поэтому он не согласуется с наблюдениями. Добавим, что в ОТО допустим не только мир положительной кривизны, но и миры с отрицательной и нулевой кривизной. В нашей механической задаче это означает, что полная энергия будет равна $+\frac12 mc^2$ или нулю. Оба эти случая гарантируют прохождение точки Эйнштейна  и значит, оставляют в силе только сценарий ``монотонного мира первого рода''. 

Почему обсуждая космологию мы опираемся на Фридмана, а не на Эйнштейна, предложившего своё решение проблемы на пять лет раньше? Потому что решение Эйнштейна было ошибочным. Оно описывало застывшую в точке Эйнштейна Вселенную, то есть, случай, когда наше пробное тело кто-то аккуратно подвесил в той самой точке, где силы притяжения и отталкивания уравновешивают друг друга, иначе говоря, положил на вершину потенциальной горы. Чтобы это вообще было возможно, надо ещё и подогнать точку Эйнштейна под значение полной энергии, то есть, связать три числа: радиус мира $R$, космологическую постоянную $\lambda$ и массу мира $M$  двумя уравнениями.  В нашем рассмотрении эти два уравнения выглядят так
$$
\beta=1,\qquad x=1,
$$
или
$$
\RSCH=\pi\RE, \qquad R=\RE.
$$
 Но каждый школьник знает, что положение на вершине является положением неустойчивого равновесия, а значит, рано или поздно тело покатится вниз. Космологическую постоянную $\lambda$, т.е. силу всемирного отторжения, Эйнштейн ввел нарочно, изменив первоначальные уравнения ОТО ради того, чтобы получить для нас мир вечный и неизменный. Но всё зря. Так интуиция иногда обманывает даже самые сильные умы. Не обманывает только математика! А космологическая постоянная спустя 80 лет всё-таки пригодилась. В конце 1990-х годов было открыто ускоренное расширение Вселенной. Значит, точка Эйнштейна в ходе эволюции мира уже пройдена. 

Нам кажется, что случай Эйнштейна ниже изображен наглядно.

\includegraphics[width=12cm,height=8cm]{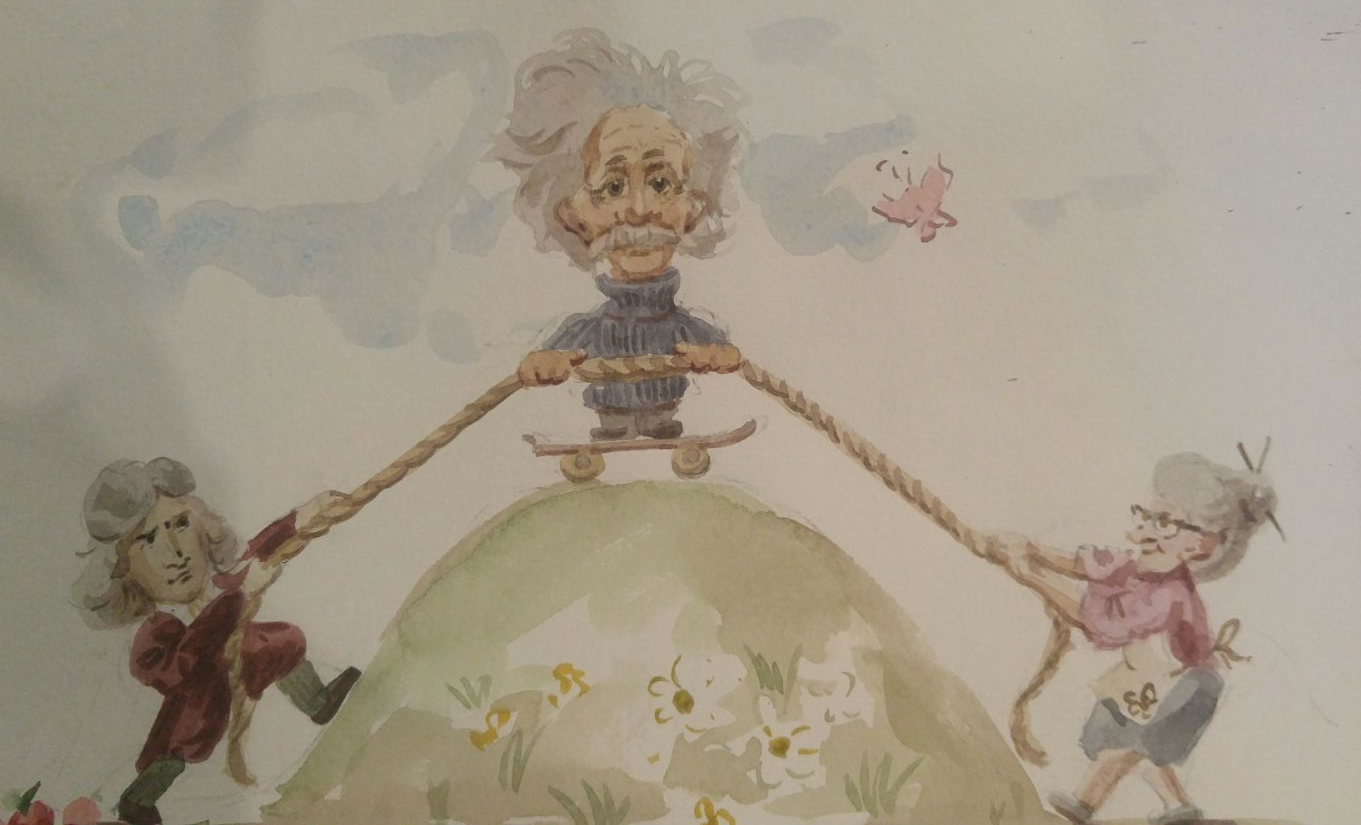}
Рис.3

\subsection*{Несколько чисел}
Спустя три месяца после открытия уравнений динамики Вселенной Фридман выпускает для широкой публики книгу ``Мир как пространство и время''. Она начинается цитатой из Козьмы Пруткова, а кончается стихами Г.Р. Державина. В конце книги Фридман пишет: ``{\it Теория Эйнштейна оправдывается на опыте; она объясняет старые, казавшиеся необъяснимыми явления и предвидит новые поразительные соотношения. Вернейший и наиболее глубокий  способ изучения, при помощи теории Эйнштейна, геометрии мира и строения нашей Вселенной состоит в применении этой теории ко всему миру и в использовании астрономических исследований. Пока этот метод немногое может дать нам... Но в этих обстоятельствах нельзя не видеть лишь затруднений временных; наши потомки, без сомнения, узнают характер Вселенной, в которой мы обречены жить}...''

Можно дополнить получившуюся картину современными оценками параметров эволюции Вселенной. При жизни Фридмана (он умер от тифа в 1925 г.) не было известно ничего. Сейчас у нас есть данные о постоянной Хаббла, о процентном отношении вкладов материи, включая невидимую нами тёмную материю, о тёмной энергии, роль которой возможно играет космологическая постоянная. 

Постоянная Хаббла $H$ названа в честь американского астронома, работавшего на крупнейшем в то время 2,5-метровом оптическом телескопе в Маунт-Вилсон и обнаружившего в 1929 г. прямую пропорциональность красного смещения света галактик расстоянию до них. В наших обозначениях $H=R'/R$, современное значение $H=2.3\times 10^{-18}$ с$^{-1}$. Тёмная энергия, которая возможно сводится к космологической постоянной, даёт примерно $70\%$ вклада в полную плотность энергии, а вещество, включающее в себя и невидимую загадочную тёмную материю, даёт примерно $30\%$. Есть ещё излучение, плотность энергии которого убывает при расширении Вселенной как $1/R^4$, но его вкладом в первом приближении мы можем пренебречь. Вклад кривизны пространства не превышает $0.4\%$, поэтому пространство обычно предполагается плоским. Мы, однако, для наглядности сохраним за переменной $R$ титул радиуса мира и будем считать пространство Вселенной 3-мерной сферой, вслед за Эйнштейном и Фридманом.  В нашей механической модели Вселенной соотношение между тёмной энергией и энергией вещества есть отношение потенциальных энергий 
$$
\frac{U_{\mbox{анти-Гук}}}{U_{\mbox{Ньютон}}}=\frac{\frac{m\omega^2R}{2}}{\frac{\alpha GmM}{R^2}}=\frac12\left(\frac{R}{\RE}\right)^3\equiv \frac{x^3}{2}\approx 2.3,
$$ 
следовательно, $x\approx\sqrt[3]{4.6}\approx 1.7$. Соотношение между вкладом кривизны (у нас это полная энергия) и вкладом вещества
$$
\frac{E}{U_{\mbox{Ньютон}}}=\frac{1}{\frac{2\beta}{3x}}=\frac{3x}{2\beta}<0.013,
$$
дает нам оценку параметра $\beta>190$.
Для безразмерной постоянной Хаббла $\bar H$, график которой мы строили при разных значениях $\beta$ получаем
$$
\bar H\equiv\frac{v}{cx}=
H\frac{\RE}{c}\approx 0.77\cdot 10^{-26} \RE,
$$
если $\RE$ выражено в метрах. 
Из нашей формулы (\ref{eq:H}) получаем
$$
\bar H(x=1.7)\approx 0.68\sqrt{\beta}>9.4,
$$
следовательно, $\RE>1.2\cdot 10^{27}$ м или в более адекватных единицах, мегапарсеках $\RE>6\cdot 10^4$ мпс.
Таким образом, из данных о постоянной Хаббла мы можем получить оценку радиуса мира $R> 1.7\RE\approx 10^5$ мпс или $R>2\cdot 10^{27}$ м.
Для массы мира получаем $M> 1.5\cdot 10^{57}$ кг. Для космологической постоянной Эйнштейна $\lambda=3\omega^2<4.3\cdot 10^{-35}$ с$^{-2}$. Для радиуса Шварцшильда Вселенной получаем $\RSCH>3.6\cdot 10^7$ мпс.

\subsection*{Рождение Вселенной из ничего?}
...``{\it является возможность также говорить о ``сотворении мира из ничего'', но всё это пока должно рассматривать как курьёзные факты, не могущие быть солидно подтверждёнными недостаточным астрономическим экспериментальным материалом}...''   (А. Фридман, ``Мир как пространство и время'').

В обеих наших моделях космологии было что-то искусственное. В примере с катанием на санях кто-то должен разогнать сани, во втором примере кто-то должен выстрелить из рогатки... Не следует ли представить всё проще? Чтобы перевалить через одну горку не лучше ли будет сначала забраться на горку более высокую и начать спуск отуда? Тогда будет понятно, откуда взялись огромная кинетическая энергия и огромная по модулю, но отрицательная, потенциальная.  Начальная полная энергия может быть очень малой, и даже нулевой. Вселенная может родиться из ничего! Приблизительно об этом говорит сценарий так называемой инфляции, когда Вселенная рождается как флуктуация вакуума и за первые $10^{-30}$ с своего существования увеличивает свои размеры не меньше, чем в $10^{30}$ раз. Вся огромная масса вещества Вселенной рождается при распаде первичного вакуума и переходе от эпохи инфляции к эпохе Фридмана. 

\includegraphics[width=8cm,height=5cm]{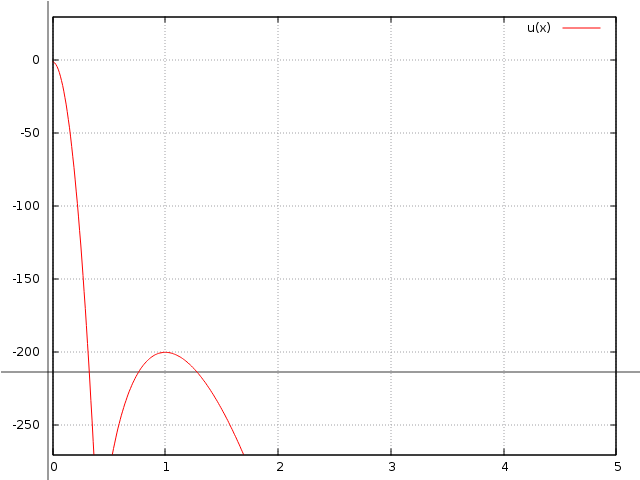}

Так ли проста космология, как здесь было представлено? Конечно, здесь было сказано далеко не всё. Вещество, которое Эйнштейн, за ним Фридман и мы с вами считали пылью, на самом деле включает в себя и излучение и ведёт себя по-разному при разных плотностях и температурах. Мы не уверены в том, что происходит в мире элементарных частиц при энергиях на много порядков выше тех, которые доступны современным ускорителям. В ранней Вселенной происходили грандиозные фазовые переходы, распад вакуума и рождение из него новых и новых частиц. Эволюция Вселенной является неравновесным и необратимым процессом. Первым физиком всерьёз обратившим на это внимание был Георгий Гамов, которому вместе с учениками, удалось вывести соотношение между количествами лёгких элементов системы Менделеева во Вселенной, предсказать существование реликтового излучения и правильно оценить его температуру, но это отдельная история.


\end{document}